\documentclass[fleqn,twoside]{article}
\usepackage{amsmath}
\usepackage{espcrc2}
\usepackage{graphicx}

\usepackage{hyperref}

\newcommand{\um}[1]{\ensuremath{\,\mathrm{#1}}} % units of measure
\newcommand{\particle}[1]{\ensuremath{\mathrm{#1}}}
\newcommand{\micro}{\ensuremath{\mu}}
\newcommand{\Ohm}{\ensuremath{\Omega}}

\title{Design and test results of the AMS RICH detector}

\author{D. Casadei\address[DC]{Physics Department, Bologna University
        and INFN Bologna, Via Irnerio 46, I-40126 Bologna, Italy.}%
        \thanks{\url{Diego.Casadei@bo.infn.it}} 
        for the AMS-RICH Collaboration
}

\begin{document}

\begin{abstract}
 The AMS-02 detector will operate for at least 3 years on the
 International Space Station, measuring cosmic ray spectra at about
 400 km above sea level over a wide range of geomagnetic latitude.
 The proximity focusing ring imaging \v{C}erenkov counter of AMS-02
 will measure the particle velocity $\beta$ with $\approx 0.1\%$
 uncertainty, making possible to discriminate Beryllium isotopes up to
 about 15 GeV/nucl.  In addition its charge measurement will allow to
 study the elemental composition of cosmic rays up to Iron.  A
 prototype of the RICH detector was tested with cosmic rays and on a
 ion beam accelerated by SPS, at CERN (October 2002).\newline
 [\emph{Talk given at the ``8th Topical Seminar on Innovative Particle
 and Radiation Detectors'', Siena (Italy), 21--24 October 2002.}]
\end{abstract}

\maketitle

\section{INTRODUCTION}

 The \emph{Alpha Magnetic Spectrometer} (AMS)~\cite{amsfirst} is a
 particle detector that will be installed on the International Space
 Station (ISS) to measure cosmic ray fluxes for at least three years.
 The AMS goals are the search for cosmic antimatter, the search for
 dark matter signatures, and the measurement of primary Cosmic Ray
 (CR) spectra below 1 TeV.

\begin{figure*}[t!]
\centering
\includegraphics[width=\textwidth]{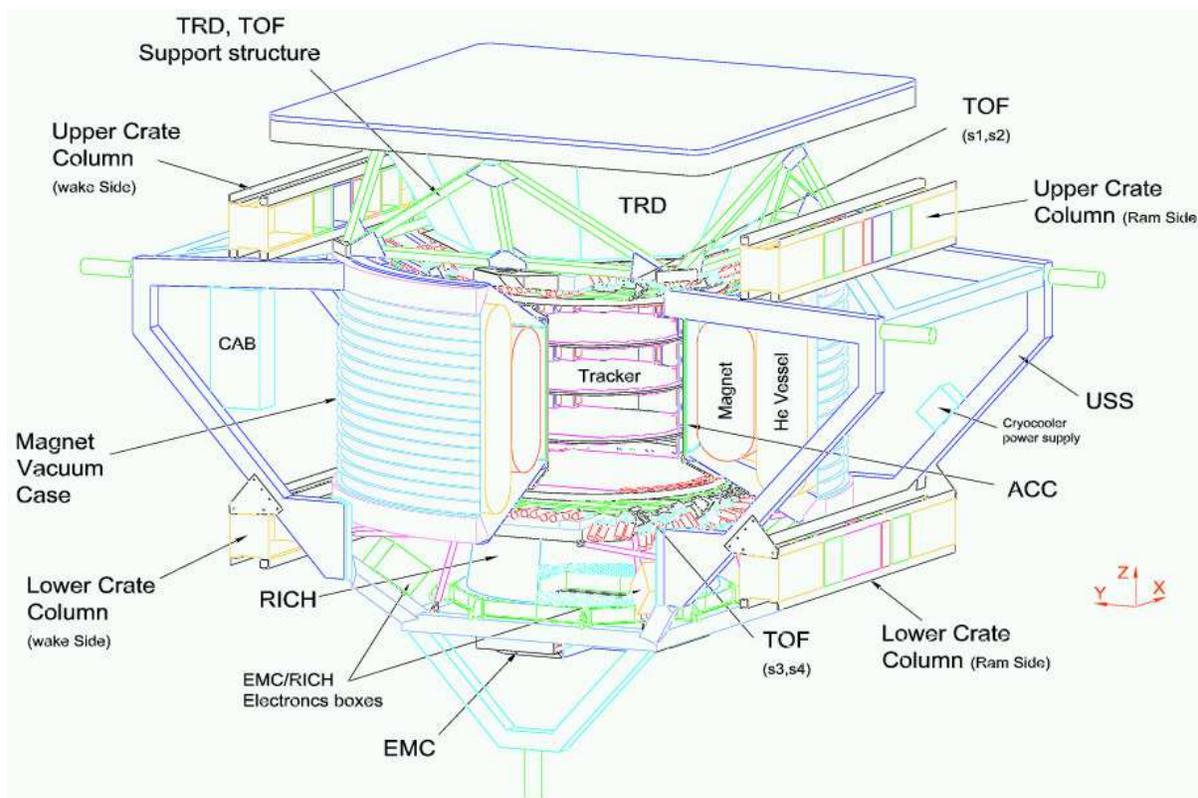}
\caption{The AMS detector to be installed on ISS (AMS-02).}\label{AMS2}
\end{figure*}

 In 1998 a first version of the detector (called AMS-01) has flown
 aboard of the space shuttle Discovery during the NASA STS-91 mission,
 and it operated successfully for ten days collecting about hundred
 million events (refer to \cite{ams1rep} for a review of the AMS-01
 results).  For the ISS 3-year mission, a new version of the detector
 (called AMS-02, figure~\ref{AMS2}) is being developed.

 The new apparatus will consist of a transition radiation detector
 (TRD) followed by two of the four scintillator planes of the time of
 flight (TOF) system, placed above the superconducting magnet
 enclosing a Silicon tracker surrounded by an anticoincidence (ACC)
 system.  Below the magnet, the other two planes of the TOF are
 followed by the proximity focusing ring imaging \v{C}erenkov counter
 (RICH) and by the electromagnetic calorimeter (EMC).

 The large acceptance of the apparatus (about 0.4 \um{m^2} sr) and the
 long duration of the mission (at least 3 years) will allow for a
 detailed study of the spectra of Hydrogen and Helium isotopes, that
 can be used to study solar modulation on a weekly basis and to check
 different CR propagation models.  In addition to very precise
 measurements of the electrons and positrons spectra, AMS will also
 collect an unprecedented amount of ions events, that are of paramount
 importance for the fine tuning of the parameters of any propagation
 model \cite{heinbach95}.  Finally, the detector will be able to
 measure high energy gamma rays up to few hundred GeV
 \cite{battiston00}.

\section{THE DESIGN OF THE AMS-02 RICH}

\begin{figure}[t!]
\centering
\includegraphics[width=\columnwidth]{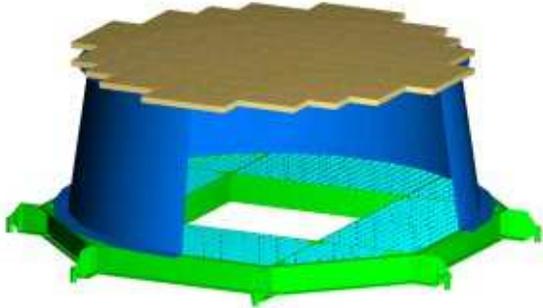}
\caption{The AMS-02 RICH subdetector.}\label{RICH}
\end{figure}

 In order to measure the particle velocity with high accuracy, a
 \v{C}erenkov detector is highly desirable.  Howewer, the need for a
 wide geometrical acceptance makes impossible the use of an ordinary
 ring imaging detector: a proximity focusing technique has to be
 adopted.  Hence the RICH subdetector of AMS-02 consists of a flat
 radiator plane, where the \v{C}erenkov photons are emitted, followed
 by a gap where these photons propagate in vacuo until they reach the
 pixel plane, parallel to the radiator (figure~\ref{RICH}).
 The hole in the middle of the pixel plane is exactly above the
 electromagnetic calorimeter, in order not to have a big amount of
 matter in front of it.

 The basics constituents of the AMS-02 RICH are the radiator plane,
 the conical mirror, and the pixel plane with 680 16-channel
 phototubes (PMT) and segmented plastic light guides (LG) to decrease
 the dead area of the pixel plane.

\subsection{The radiator plane}

 Particle mass identification depends on the momentum and velocity
 resolution of the detector:
\begin{equation}
 \frac{\Delta m}{m} = \frac{\Delta p}{p} \oplus \gamma^2
                      \frac{\Delta\beta}{\beta}
\end{equation}
 (where $\oplus$ denotes the sum in quadrature).  Given the good
 momentum resolution of the AMS-02 tracker, it comes out that particle
 identification with the RICH can be achieved over a range
 $p_\text{min}/p_\text{max} = 3 \div 4$ \cite{buenerd00}.

 The interesting range for constraining the parameters of propagation
 models is roughly $(1 \div 15)$ GeV/nucl, where the ratio between
 Lithium, Beryllium, Boron and Carbon and the ratio between sub-Fe
 elements and Iron are expecially useful to test different models
 \cite{heinbach95}.  Howewer, no single radiator can be used to cover
 this whole range, hence it is necessary to make a choice.  Due to the
 fact that the balloon measurements collected events principally in
 the lower energy range, the AMS Collaboration agreed that the RICH
 has to focus on the higher range.

\begin{figure}[t]
\centering
\includegraphics[width=\columnwidth]{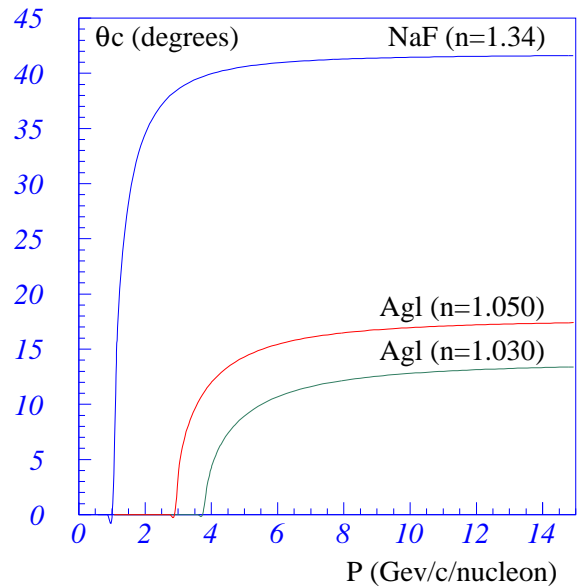}
\caption{\v{C}erenkov angle as function of momentum per nucleon, for
Sodium Fluoride and aerogel.}\label{angles}
\end{figure}

 The only solid material with refractive index small enough to allow
 particle discrimination up to 15 GeV/$c$ per nucleon is aerogel, that
 is available with refractive indexes of 1.03 and 1.05, and has a
 maximum \v{C}erenkov angle of about 15 degrees (figure~\ref{angles}).

\begin{figure}[t]
\centering
\includegraphics[width=\columnwidth]{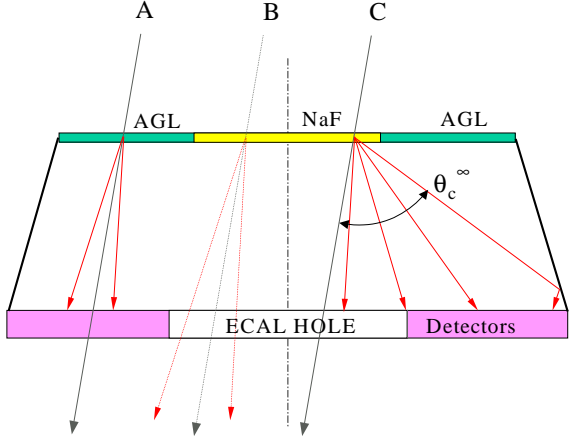}
\caption{Possible configuration fo the radiator plane, where a central
small part of NaF is surrounded by aerogel tiles.}\label{miss}
\end{figure}

 Due to the small \v{C}erenkov angles obtained with aerogel, a
 fraction of the particles crossing the radiator in the middle and
 entering the EMC hole would not be detected by the RICH (track B in
 figure~\ref{miss}).  The Collaboration has still the option to put a
 small tile of NaF as central radiator: thanks to the wider emission
 angle of NaF, this would allow for an increase of the effective
 geometric factor of order 10\%.

 Due to the greater refractive index of Sodium Fluoride, the number of
 emitted photons is greater than that of aerogel.  Hence the width of
 the radiator would be different: 3 cm for aerogel tiles, 0.5 cm for
 the NaF core.

\subsection{Conical mirror and support structure}

 The incident angle of charged particles crossing the AMS-02 RICH
 radiator has a broad distribution (roughly 40 degrees wide), and many
 events would produce photons not detectable by the pixel plane.  For
 this reason, a conical mirror is foreseen, that will increase the
 acceptance of about 30\%.

 The mirror construction is quite elaborate: first, a conical Aluminum
 mandrel is done whose external surface is lathe machined down to
 RMS roughness of order 200 nm.  Then this surface will be chemically
 plated with 60 \micro{m} of Nickel and polished again down to about
 10 nm RMS.  Finally the mandrel will serve as the ``negative'' of the
 mirror, that will be made of a carbon fiber structure 1 mm thick
 whose internal surface will be plated with Aluminum and finally
 with a protection layer of quartz or Manganese Fluoride.

 While the radiator support structure (in carbon fiber) is fixed to
 the TOF mechanics, the mirror and the pixel plane are bound together,
 with their Aluminum mechanics directly attached to the AMS-02 support
 structure.  Hence, the vibration tests and the finite element method
 simulations for the radiator are shared with those of the TOF
 subsystem, while the pixel plane and the mirror have a parallel
 vibration study.  This effort is necessary to follow all NASA
 specifications, and (both for RICH and TOF) it is carried on in
 collaboration with the Italian company Carlo Gavazzi Space SpA.

\subsection{The phototubes}

 The pixel plane of the AMS-02 RICH consists of 680 multi-channel
 Hamamatsu phototubes derived from model R5900.  This ``venetian
 blind'' R7600-00-M16 model has the photo-cathode divided into
 $4\times4$ cells, followed by 12 metal channel dynodes and a
 segmented anode with 16 read-out channels.

 The photo-cathode cells are $4.5 \times 4.5$ \um{mm^2} wide and the
 PMT is roughly a square with $(25.7 \pm 0.5)$ mm side and $(20.1 \pm
 0.5)$ mm body height, with 7 mm pins.  The voltage divider has total
 impedance 80 M\Ohm\ and the repartition (2.4 : 2.4 : 1 : 1 : 1 : 1 :
 1 : 1 : 1 : 1 : 1.2 : 2.4 : 2.4) was chosen in order to achieve a
 good linearity.  Such PMTs have a gain of order $2 \times 10^6$ at
 voltages in the range $(770 \div 870)$ V and their pixel cross-talk
 is usually below 5\%.

\begin{figure}[t]
\centering
\includegraphics[width=\columnwidth]{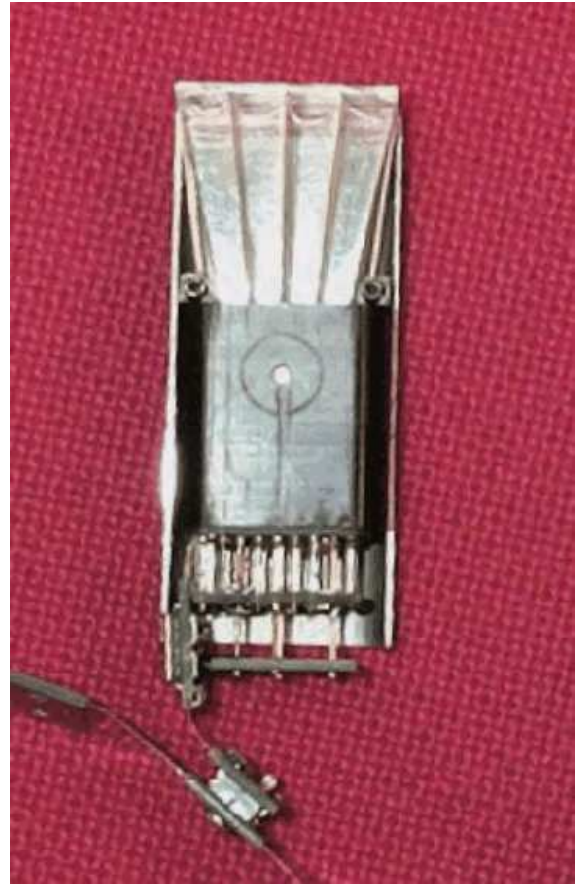}
\caption{A picture of the PMT assembly.}\label{pmt}
\end{figure}

 Figure~\ref{pmt} shows a picture of the RICH phototube assembly: the
 PMT is enclosed by the black housing, to which the plastic light
 guides are bound with Kevlar wires.  The printed circuit boards below
 the PMT contain the voltage divider and the anodes read-out circuits,
 connected to an amplifier.  Signals from several PMTs are read via a
 flat cable that is visible in the lowest part of the picture.
 Finally, half of the enclosing magnetic shielding is shown.
  
\begin{figure}[t!]
\centering
\includegraphics[width=\columnwidth]{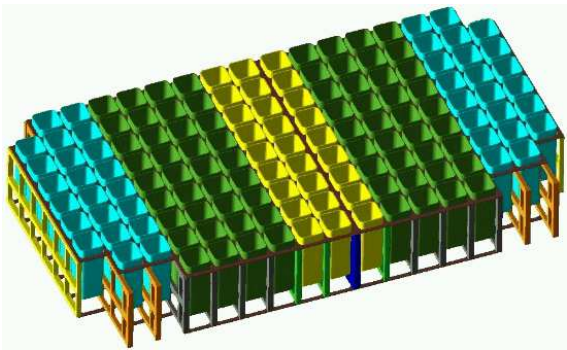}
\caption{The shielding set along the $x$ direction is made of boxes of
different thickness, increasing toward the middle.}\label{grid}
\end{figure}

 The PMT were tested in magnetic field, with and without magnetic
 shielding, in order to understand their behavior in the final
 configuration, and it comes out that the gain will change at most
 20\% in the worst cases (with $B \approx 300$ G along the AMS $x$
 direction).  In particular, the shielding boxes along the $x$
 direction will have different thickness (from 6.0 mm to 1.2 mm), in
 order to compensate for collective effects (figure~\ref{grid}).

\section{PROTOTYPE TEST}

 A prototype for the RICH radiator was built at ISN Grenoble, and used
 to test different radiators using cosmic ray and beam particles.  The
 prototype (radiator, pixel plane and front-end electronics) is placed
 inside a vacuum chamber, to avoid environmental light and to ensure
 that the refractive index between the radiator and the pixel plane is
 the same as during the ISS operation.

\begin{figure}
\centering
\includegraphics[width=0.7\columnwidth]{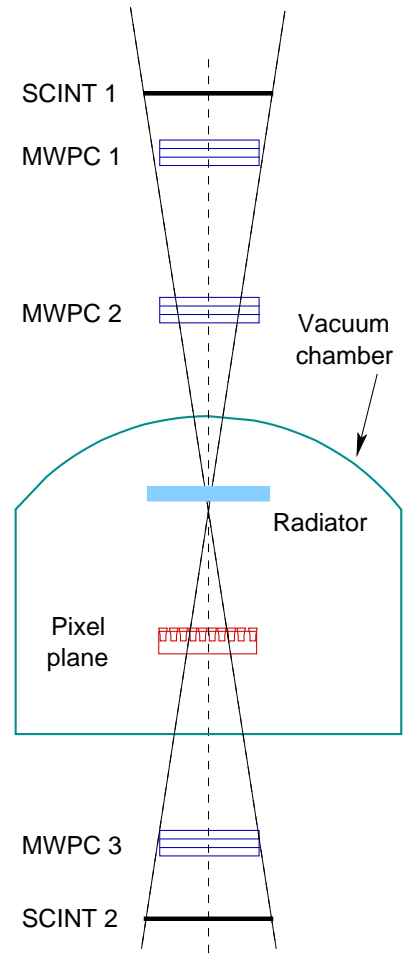}
\caption{RICH prototype setup for cosmic rays.}\label{rich-cr}
\end{figure}

 Figure \ref{rich-cr} shows the setup used for CR runs: the charged
 particles that cross both scintillators are tracked by three
 multi-wire proportional chambers (MWPC), placed above and below the
 prototype.  For beam runs, there are only one MWPC and a couple of
 thin small scintillators in front of the vacuum chamber, in order to
 minimize the amount of material placed before the detector.

\begin{figure}
\centering
\includegraphics[width=\columnwidth]{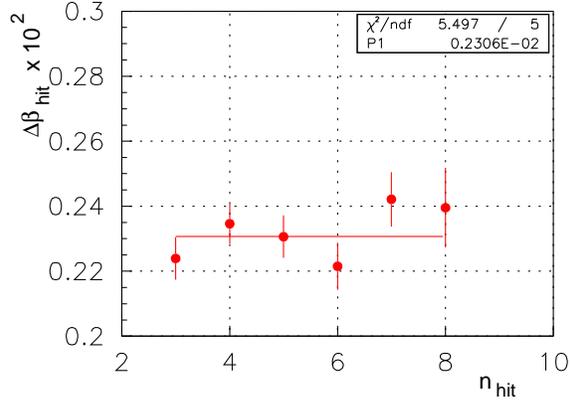}
\caption{$\beta$ resolution for one CR muon.}\label{rich-cr-beta}
\end{figure}

\begin{figure}
\centering
\includegraphics[width=\columnwidth]{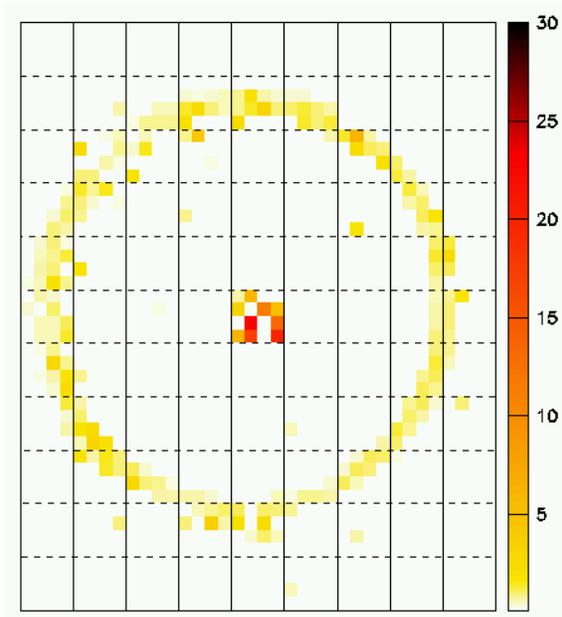}
\caption{Li event at CERN SPS, Oct.\ 2002.}\label{ring}
\end{figure}

 Figure \ref{rich-cr-beta} shows the $\beta$ resolution for a given CR
 event, taking into consideration all used pixels (every hit gives an
 independent estimate of the \v{C}erenkov angle).  On the other hand,
 figure~\ref{ring} shows the hits configuration for a Lithium ion
 during the beam test carried on at CERN during October 2002.  During
 this test the Pb beam accelerated by the SPS was directed on a Be
 target, and the 20 GeV/nucl fragments were selected by requiring
 different values for the ratio $A/Z$.  The RICH prototype, together
 with prototypes of the AMS-02 TOF and tracker, collected more than 10
 million events with $A/Z = 2$, 3/2 (\particle{^3He}), 7/4
 (\particle{^7Be}), and 1 (protons), and the data analysis is in
 progress.

\end{document}